\documentclass[aps,prl,twocolumn,showpacs,superscriptaddress,preprintnumbers,amsmath,amssymb]{revtex4-1}

\usepackage{graphicx}
\usepackage{epsfig}
\usepackage{dcolumn}
\usepackage{bm}
\usepackage{longtable}
\usepackage{color}

\begin{document}


\title{Disorder-dependent superconducting phase-diagram at high magnetic fields in Fe$_{1 + y}$Se$_{x}$Te$_{1-x}$ ($x \sim 0.4$)}



\author{T.\ Gebre}
\affiliation{National High Magnetic Field Laboratory, Florida
State University, Tallahassee-FL 32310, USA}
\author{G.\ Li}
\affiliation{National High Magnetic Field Laboratory, Florida
State University, Tallahassee-FL 32310, USA}
\author{J.\ B.\ Whalen}
\affiliation{National High Magnetic Field Laboratory, Florida
State University, Tallahassee-FL 32310, USA}
\affiliation{Department of Chemical and Biomedical Engineering, Florida State
University, Tallahassee-FL 32310, USA}
\author{B.\ S.\ Conner}
\affiliation{National High Magnetic Field Laboratory, Florida
State University, Tallahassee-FL 32310, USA}
\author{H.\ D.\ Zhou}
\affiliation{National High Magnetic Field Laboratory, Florida
State University, Tallahassee-FL 32310, USA}
\author{G.\ Grissonanche}
\affiliation{National High Magnetic Field Laboratory, Florida
State University, Tallahassee-FL 32310, USA}
\author{M.\ K.\ Kostov}
\affiliation{National High Magnetic Field Laboratory, Florida
State University, Tallahassee-FL 32310, USA}
\affiliation{Department of Chemical and Biomedical Engineering, Florida State
University, Tallahassee-FL 32310, USA}
\author{A.\ Gurevich}
\affiliation{National High Magnetic Field Laboratory, Florida
State University, Tallahassee-FL 32310, USA}
\affiliation{Physics Department, Old Dominion University, Norfolk-VA 23529, USA}
\author{T.\ Siegrist}
\affiliation{National High Magnetic Field Laboratory, Florida
State University, Tallahassee-FL 32310, USA}
\affiliation{Department of Chemical and Biomedical Engineering, Florida State
University, Tallahassee-FL 32310, USA}
\author{L.\ Balicas} \email{balicas@magnet.fsu.edu}
\affiliation{National High Magnetic Field Laboratory, Florida
State University, Tallahassee-FL 32310, USA}


\date{\today}

\begin{abstract}
We compare the superconducting phase-diagram under high magnetic
fields (up to $H = 45$ T) of Fe$_{1+y}$Se$_{0.4}$Te$_{0.6}$ single
crystals originally grown by the Bridgman-Stockbarger (BRST)
technique, which were annealed to display narrow superconducting
transitions and the optimal transition temperature $T_c \gtrsim 14$
K, with the diagram for samples of similar stoichiometry grown by
the traveling-solvent floating-zone technique as well as with the
phase-diagram reported for crystals grown by a self-flux method. We
find that the so-annealed samples tend to display higher ratios
$H_{c2}/T_c$, particularly for fields applied along the inter-planar
direction, where the upper critical field $H_{c2}(T)$ exhibits a
pronounced downward curvature followed by saturation at lower
temperatures $T$. This last observation is consistent with previous
studies indicating that this system is Pauli limited. An analysis of
our $H_{c2}(T)$ data using a multiband theory suggests the emergence
of the Fulde-Ferrel-Larkin-Ovchnikov state at low temperatures. A
detailed structural x-ray analysis, reveals no impurity phases but
an appreciable degree of mosaicity in as-grown BRST single-crystals
which remains unaffected by the annealing process. Energy-dispersive
x-ray analysis showed that the annealed samples have a more
homogeneous stoichiometric distribution of both Fe and Se with
virtually the same content of interstitial Fe as the non-annealed
ones. Thus, we conclude that the excess of Fe, in contrast to
structural disorder, contributes to decrease the superconducting upper-critical fields of this series.
Finally, a scaling analysis of the fluctuation conductivity in the
superconducting critical regime, suggests that the superconducting
fluctuations have a two-dimensional character in this system.
\end{abstract}


\maketitle

\section{Introduction}

The Fe$_{1+y}$(Te,Se) series is a quite unique superconducting
system: its end member, FeSe is a superconductor with a
superconducting transition temperature $T_c \simeq 8$ K, \cite{hsu}
which can be increased up to $\sim 14$ K with the partial
substitution of Se for Te, \cite{fang1,yeh} or increased all the way
up to 36.7 K by the application of hydrostatic
pressure.\cite{medvedev} The other end member of this series
Fe$_{1+y}$Te is not superconducting and instead exhibits a
simultaneous structural and magnetic phase transition from
tetragonal to monoclinic lattice accompanied by antiferromagnetism
(AFM) near $T_N \simeq 60$ to $70$ K (Refs.
\onlinecite{fruchart,bao, li}).  This AFM structure is distinct from
those seen in the undoped FeAs based compounds.\cite{cruz, huang,
zhao}  The AFM order in Fe$_{1+y}$Te propagates along the diagonal
direction of the original undistorted Fe square lattice
\cite{fruchart, li}, while in FeAs the wave-vector of the spin density wave (SDW) propagates
along the nearest neighbor direction of the original Fe square
lattice. \cite{cruz, huang, zhao}

Density functional theory (DFT) calculations indicate
that the Fermi surfaces (FS) of both FeSe and FeTe are similar to
those of the other FeAs based compounds, consequently they should
satisfy the nesting condition for the wavevector $Q_{\pi,\pi} =
(\pi,\pi) $. \cite{subedi} This apparent lack of universality would
seem to question the exclusive role of the interband pairing via
$Q_{\pi,\pi}$ magnetic fluctuations in the pairing mechanism for all
the iron-based superconductors. However, the observation of a spin
resonance below $T_c$ together with an enhancement of the spin
fluctuations near it, indeed suggests a spin-fluctuations
mediated superconducting pairing mechanism for this
system.\cite{qiu,imai} Initial angle resolved photoemission (ARPES) measurements in the
Fe$_{1+y}$Te revealed a pair of nearly compensated
electron-hole FS pockets, with no evidence for the FS nesting
instability associated with $Q_{\pi,\pi}$. \cite{hasan} This could
suggest the possibility of a distinct mechanism for the
superconductivity and the magnetic-order in the Fe chalcogenides
when compared to the Fe arsenides. But a recent ARPES study on
superconducting Fe$_{1.03}$Te$_{0.7}$Se$_{0.3}$ reveals a holelike
and an electronlike FS located at the center and the corner of the
Brillouin zone, respectively. These FSs are nearly nested for
$Q_{\pi,\pi}$. \cite{nakayama} The same study
reports an isotropic superconducting gap along the holelike FS with
a gap $\Delta$ of $ \sim 4$ meV $(2 \Delta = k_B T_c \sim 7)$, thus
providing evidence for strong-coupling superconductivity.
But, this is in contrast with a recent angle-dependent heat
capacity study in Fe$_{1+y}$Se$_{0.45}$Te$_{0.55}$
which finds evidence for a significant gap anisotropy on the
electron pockets. \cite{wen} Superconductivity in Fe$_{1+y}$Se was shown to be extremely sensitive to stoichiometry. \cite{mcqueen}
In the Fe$_{1+y}$(Te,Se) series, the crystal structure resembles that of the
iron arsenides \cite{hsu} with the Fe square planar sheets
forming from the edge-sharing iron chalcogen tetrahedral
network.  But it also allows the partial occupation of iron in the interstitial sites of the (Te, Se) layers, resulting
in a \emph{nonstoichiometric} composition for the Fe$_{1+y}$(Te,Se) series, where $y$
represents the excess Fe at the interstitial sites. \cite{fruchart,bao} It remains unclear
how the geometry of the FS evolves with the incorporation of interstitial Fe, but it is claimed to
suppress superconductivity. \cite{mao}

One remarkable feature of these compounds is their extremely large upper critical fields $H_{c2}$.
For instance, according to Ref. \onlinecite{petrovic} for a
Fe$_{1.05}$Te$_{0.89}$Se$_{0.11}$ single-crystal with a middle point
$T_c$ of just 11 K, one observes an $H_{c2} (T \rightarrow 0 \text{
K}) $ of $\sim 35$ T. Or for a single-crystal of
Fe$_{1.11}$Te$_{0.6}$Se$_{0.4}$ with a $T_c$ close to its optimum
value of 14 K one obtains $H_{c2} (T \rightarrow 0 \text{ K}) $ of
$\sim 40$ to 45 T. \cite{fang2} Compare these values with those of,
for instance, MgB$_2$ single crystals:  $T_c = 39$ K with
$H_{c2}^{c}(0) = 3.5 $ T and $H_{c2}^{ab}(0) = 17$ T, for fields
along the c-axis and along a planar direction, respectively.
\cite{kwok} In fact, to achieve in MgB$_2$ upper-critical fields as
large as those observed in the Fe$_{1+y}$(Te,Se) series, strong
impurity scattering is introduced to optimize the relative
strengths of intraband scattering in $\sigma$ and $\pi$ bands of
MgB$_2$. \cite{gurevich,ag} The significant differences between
these two multi-band superconducting families are attributable to
the fact that the Fe$_{1+y}$(Te,Se) superconductors are mostly Pauli
limited \cite{petrovic} while the $H_{c2}$s in MgB$_2$
are mostly determined by the orbital pair-breaking effect. However, the
shape of the $H_{c2}(T)$ curves for Fe$_{1+y}$(Te,Se) series
measured in Refs. \onlinecite{fang2} and \onlinecite{petrovic} are
quite distinct, with the former presenting a nearly linear
dependence of $H_{c2}(T)$ for ${\bf H}\|c$ and the second displaying
a pronounced concave down curvature followed by saturation at lower
temperatures (as expected for the Pauli limiting effect). Such a
curvature has also been reported in Ref. \onlinecite{khkim}.

The Pauli-limited behavior of $H_{c2}(T)$ for Fe-chalcogenides
results from the fact that they are semi-metals with rather low
carrier density and low Fermi energies $\sim 20-50$ meV for
different FS pockets \cite{hasan,nakayama}, unlike the conventional
metallic superconductor MgB$_2$. As a result, chalcogenides have
very short coherence lengths $\xi_0\simeq \hbar v_F/2\pi T_c<$ 1 nm
and thus extremely high orbitally-limited $H_{c2}^{orb} \sim
\phi_0/2\pi\xi^2$ where $v_F$ is the Fermi velocity in the ab plane,
and $\phi_0$ is the flux quantum. This in turn greatly enhances the
Pauli effects quantified by the Maki parameter $\alpha_M =
\sqrt{2}H_{c2}^{orb}/H_p$ where $H_p [T]=1.84T_c [K]$ is the
Clogston paramagnetic limit \cite{fflo}. Chalcogenides typically
have $\alpha_M\gtrsim 1$ above the critical value at which the FFLO
instability develops. Moreover, because of their very short values
of $\xi_0$, chalcogenides are naturally in the clean limit $\ell \gg
\xi_0$ which is one of the conditions of the FFLO instability, where
$\ell$ is the mean free path due to elastic impurity scattering
\cite{fflo}. These features of chalcogenides make them good
candidates to study exotic effects at high magnetic fields, in
particular, the interplay of orbital and paramagnetic pairbreaking
for multiband pairing and the FFLO state at low temperatures. The
small Fermi energies also make superconducting properties very
sensitive to the doping level, allowing one to tune $H_{c2}$ by
small shifts of the chemical potential.

In contrast to other reports emphasizing on the details concerning the synthesis of single crystals \cite{lin},
in this work we compare the high-field phase-diagram of the
Fe$_{1+y}$(Te,Se) series, particularly the diagram for the optimally
doped compound Fe$_{1+y}$Te$_{1-x \approx 0.6}$Se$_{x\approx 0.4}$
synthesized by two methods.  The first one is based on a
traveling-solvent floating zone growth technique (TSFZ), which leads
to crystals of acceptable crystallinity displaying ``non-metallic"
resistivity, optimal $T_c$s, and transition widths $\Delta T_c
\simeq 1$ to 3 K. We compare their behavior with crystals resulting
from the Bridgman-Stockbarger (BRST) technique which in our case leads to crystals of poorer
crystallinity (i.e. larger mosaicity), wider transitions and
frequently to non-optimal $T_c$s in as grown crystals, as previously
reported by other groups. \cite{mizuguchi} These last
single-crystals were subjected to an annealing procedure which lead to
metallic resistivity, considerably sharper resistive transitions
$\Delta T_c \simeq 1$ K, and to superconducting transitions
comparable to those reported in the literature for high quality
BRST-grown single crystals as measured by SQUID magnetometry.
\cite{sales} The annealed crystals having a clear metallic behavior
preceding superconductivity, display considerably higher
upper-critical fields when compared to those showing a poor
metallicity, particularly for magnetic-fields applied along the
inter-layer direction. This increase leads to a distinct shape of
the $H_{c2}^c(T)$, i.e. from a approximately linear in $T$ behavior
as reported in Ref. \onlinecite{fang2}, to the marked concave-down
curvature followed by saturation at lower temperatures as seen in
Ref. \onlinecite{khkim}.The analysis of our experimental data using
a multiband theory of $H_{c2}(T)$ \cite{Gurevich2010} shows that
$H_{c2}(T)$ is indeed strongly Pauli limited $\alpha_M\simeq 7-10$,
predicting the FFLO state below $\approx 5$ K. A detailed single
crystal x-ray analysis reveals that the annealing process does not
affect the crystallinity or the degree of mosaicity of our single
crystals. Nevertheless, our dispersive x-ray analysis indicates that
the annealing process leads to a more uniform distribution of
interstitial Fe. Annealing yields a considerably more isotropic
phase diagram which combined with the enhancement in
$H_{c2}^{c}(T)$, suggests that the variations in the content of
interstitial Fe contributes to the suppression of superconductivity.
Finally, we show, through a scaling analysis of the fluctuation
conductivity in the neighborhood of the superconducting transition,
that the superconducting fluctuations in this system are
two-dimensional in character.

\section{Sample preparation}

Single crystals of Fe$_{1+x}$Te$_{1-y}$Se$_y$, $0.05 \leq x \leq 1$
and $0.1 \leq y \leq 0.5$ were grown by using the traveling-solvent
floating-zone growth technique (TSFZ) and the Bridgman-Stockbarger
(BRST) techniques. Starting materials with nominal purities of 4N for
Fe and 5N for Te and Se were used. These were handled in an
argon-filled glove box which kept the oxygen content below 1ppm.
Mixtures of Fe, Te, and Se were ground and pelletized and sealed in
quartz ampoules under vacuum and heated at 400 $^{\circ}$C for 24 h.
The reacted material was reground and doubly sealed in two quartz
ampoules. For samples grown through the TSFZ technique, the doubly
sealed quartz ampoules were loaded into an optical floating-zone
furnace (NEC Nichiden Machinery 15HD), equipped with two 1500 W
halogen lamps. The ampoule was rotated at 20 rpm, and translated at
a rate of 1 to 2 mm/h. The as-grown samples were annealed by heating
to 800 $^{\circ}$C for 48 h, and then cooled at a rate of 100
$^{\circ}$C/h to 420 $^{\circ}$C and held at this temperature for
additional 30 h, followed by the cooling of the furnace to room
temperature.

A double wall quartz ampoule was also used in the case of crystals
grown by the BRST technique.  The inner ampoule was narrowly tipped at
its bottom and sealed at the top. This Bridgman ampoule was
inserting into a three-zone temperature gradient vertical furnace
and lowered at a speed of 4 mm/h.  The growth temperature was
decreased at a rate of 3 $^{\circ}$C/h down to 440 $^{\circ}$C and
subsequently quenched to room temperature. Some of the crystals as
well as Se powder were placed into two different quartz crucibles,
and both placed inside an evacuated quartz ampoule which was sealed
under vacuum and heated slowly to 400 $^{\circ}$C and then annealed
for ten days.

\begin{figure}[ht]
\begin{center}
\epsfig{file=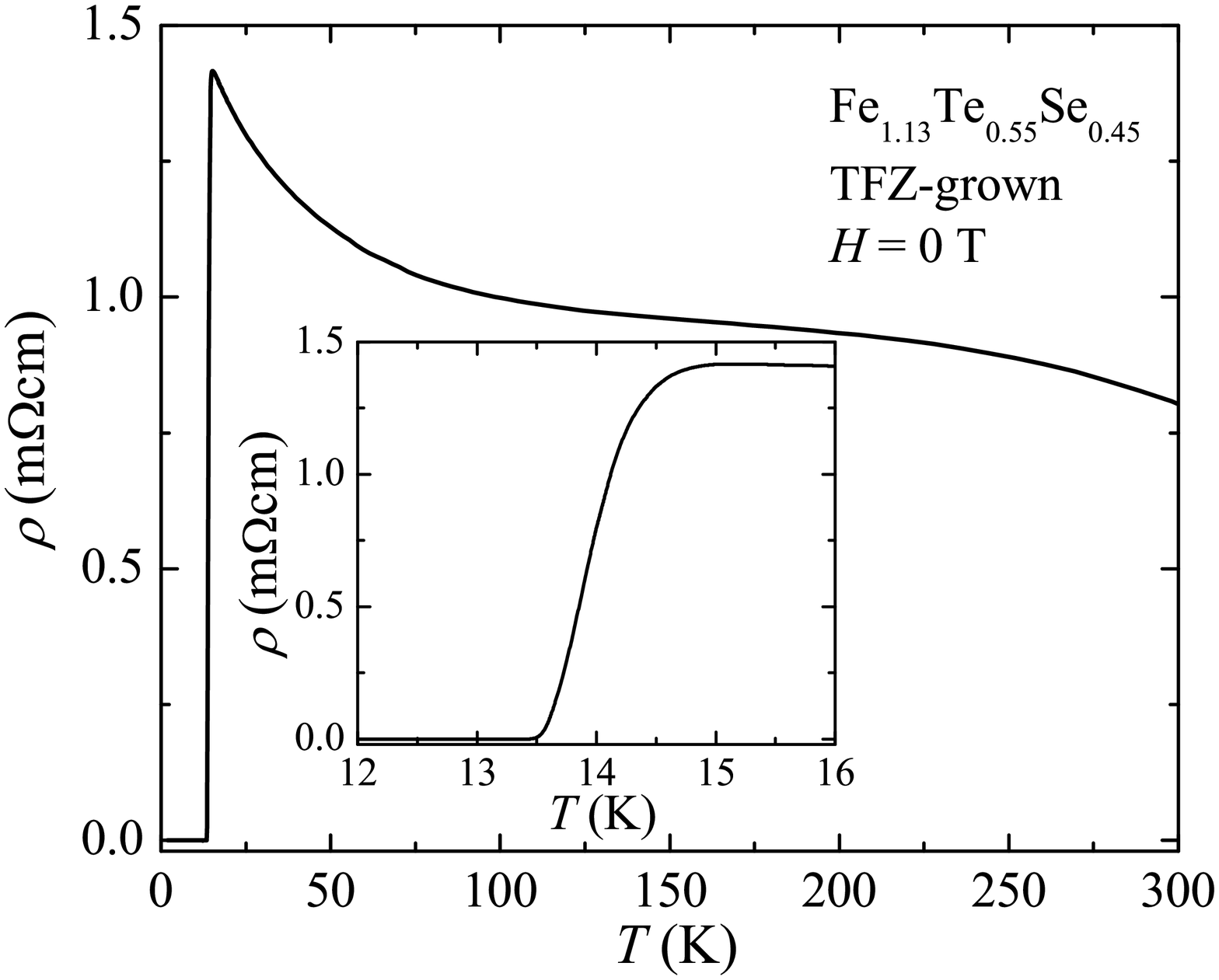, width = 8.6 cm}
\caption {(color online) Top panel: Resistivity as a function of temperature for a Fe$_{1.13}$Se$_{0.45}$Te$_{0.55}$ single crystal synthesized by the traveling floating zone method under zero magnetic field. Inset: Resistive superconducting transition under zero-field in a limited temperature range.}
\end{center}
\end{figure}
The so-obtained crystals were checked by both single-crystal and
powder x-ray diffraction (XRD) techniques. From as-grown single
crystals, shards were cleaved to obtain suitable samples of
approximately 0.3 mm $\times$ 0.3 mm $\times$ 0.03 mm.  An
Oxford-Diffraction Xcalibur-2 CCD diffractometer with Mo K$_{\alpha}$  source
was used to collect reflections.  CrysAlis Pro 171.33.55 software
(Oxford Diffraction) was used for the unit cell refinement
and analytical absorption correction.  Final structure refinements
were conducted using SHELXTL.  For the powder diffraction runs, the
single-crystals were ground by hand for approximately 1 minute using
an agate mortar and pestle.   Patterns were detected by a
Huber-Guinier 620 camera with a Rigaku Ultrax Cu K$_{\alpha}$  direct drive
rotating anode source.  WinPow software was used for unit cell parameter
least-squares calculations and peak fitting.
Further details are provided in Ref. \onlinecite{ICSD}.

Scanning electron-microscope energy-dispersive x-ray analysis
(SEM-EDX) was performed by using a JEOL 5900 scanning electron
microscope (30 kV accelerating voltage) equipped with IXRF energy
dispersion spectroscopy software (IXRF Systems, Inc.) in order to
determine the elemental compositions of the single crystals.
We found that this EDX set-up would show a tendency
to underestimate by nearly 3 \% the overall fraction of Fe when
comparing with detailed single-crystal x-ray refinements.
We must clarify that the EDX set-up was not calibrated against an absolute standard.

Longitudinal resistivity was measured using a conventional
four-contact method in continuous magnetic fields up to 45 T by
using either a PPMS or the hybrid magnet of the National High
Magnetic Field Laboratory in Tallahassee.

\section{Sample Characterization}

\begin{figure}[htb]
\begin{center}
\epsfig{file=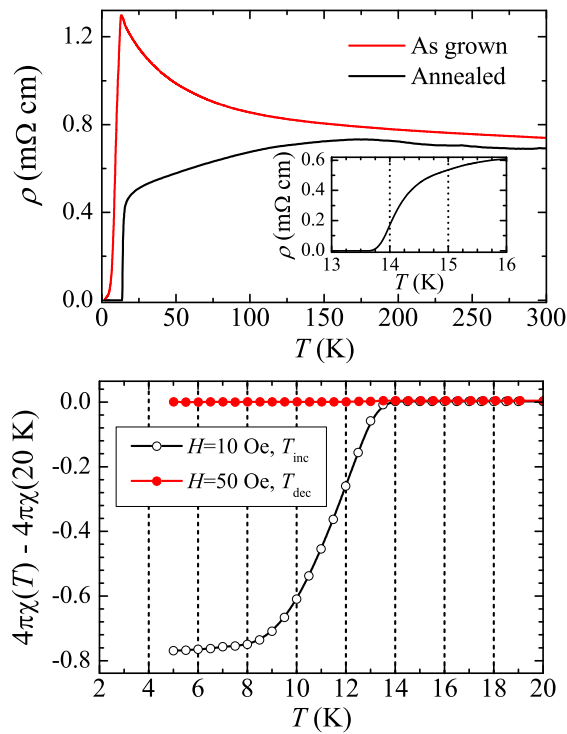, width = 7.4 cm}
\caption {(color online) Top panel: Resistivity as a function of temperature for an as-grown Fe$_{1.11}$Se$_{0.4}$Te$_{0.6}$ single crystal, synthesized through the Bridgman-Stockbarger method (red curve) and under zero magnetic field. After annealing, the sample no longer displays the weak-localization like temperature dependence. Instead it displays a metallic character (black trace). Inset: Superconducting resistive transition for the annealed crystal under zero field and in a limited temperature range. Bottom panel: Magnetic susceptibility as a function of $T$ for the same annealed single-crystal and respectively, for magnetic field-cooled (red markers) and zero-field cooled conditions (black markers).}
\end{center}
\end{figure}

Figure 1 shows a typical trace of resistivity $\rho$ as a function
of the temperature for a single crystal grown by the
traveling-solvent floating-zone method described above, whose
stoichiometry was extracted by a detailed single crystal x-ray
analysis refinement. Although the width $\Delta T_c \sim 1$ K of the
superconducting transition (see inset) is relatively narrow when
compared with the width of the transitions shown for instance in
Ref. \cite{petrovic}, the behavior of the metallic state shows the
typical negative slope $\partial \rho /
\partial T < 0 $ or logarithmic divergence of the resistivity, which
was attributed to the high content of interstitial Fe and which
leads to weak localization. \cite{mao} The same characteristic
temperature dependent resistive behavior is observed also in
as-grown samples synthesized by the Bridgman-Stockbarger method (red
line), as seen in the top panel of Fig. 2. As seen, the as-grown
samples tend to exhibit lower superconducting transition
temperatures and broader transitions. The annealing procedure
described above leads to a slightly smaller resistivity at room
temperature (black line) but most importantly to a quite distinct
temperature dependence for the resistivity, which below $T \sim 175$
K exhibits a clear metallic dependence with $\partial \rho /
\partial T > 0 $. The magnetic susceptibility on the other hand,
shows a very clear diamagnetic signal (black markers) below the
onset of $T_c$ at $T = 14$ K, as seen in the lower panel of Fig. 2.
At higher temperatures, the magnetic susceptibility is virtually
temperature independent, as reported in Ref. \onlinecite{yang} , and
only at low temperatures one observes a very mild upturn in the
susceptibility suggesting the presence of a small amount of
localized magnetic moments. Although none of the susceptibility
traces in Ref. \onlinecite{yang} show a clear diamagnetic signal
below $T_c \simeq 14 $ K but below $\sim 10$ K. \vspace{0.7 cm}
\begin{figure}[htb]
\begin{center}
\epsfig{file=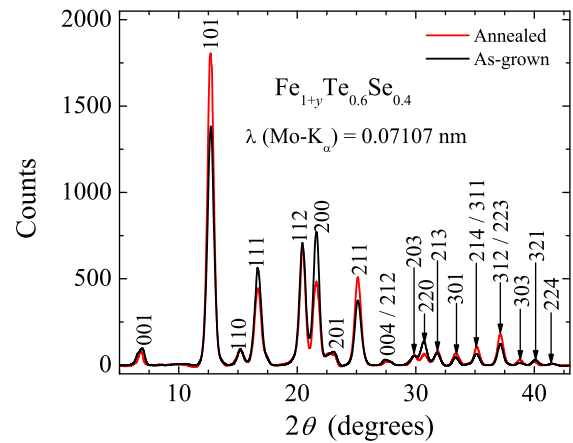, width = 7.4 cm}
\caption {(color online) Powder x-ray diffraction pattern for two Fe$_{1.11}$Se$_{0.4}$Te$_{0.6}$ single crystals grown through the Bridgman technique, collected by using the Mo-K$_\alpha$ line. Black line corresponds to the spectrum of an as-grown crystal, while the red line is the spectrum obtained from a subsequently annealed single-crystal. In both cases well-defined Bragg peaks are observed and indexed within the space-group \emph{P}4/\emph{nmm}.  We find no evidence for impurity phases or a change in structure/stoichiometry induced by the annealing process.}
\end{center}
\end{figure}
\begin{figure}[htb]
\begin{center}
\epsfig{file=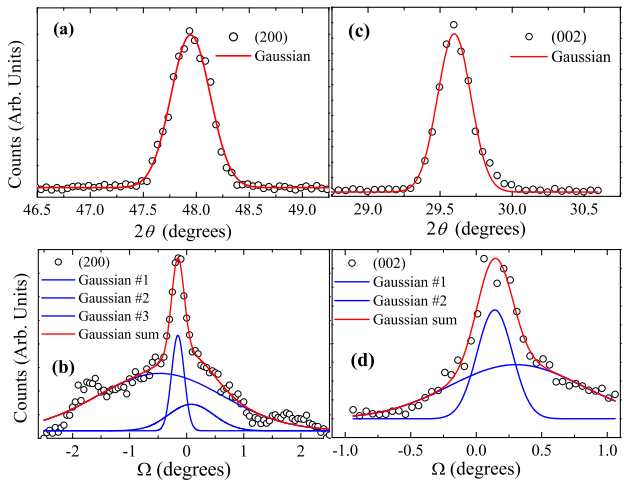, width = 8.2 cm}
\caption {(color online) (a) X-ray intensity for an as-grown BS single-crystal as a function of $2\theta$ for an angular scan around the Bragg-peak observed along the real-space direction (\emph{abc}) = (200). Red line is a fit to a Gaussian function which yields a full-width at half-maximum of $\sim 0.5^{\circ}$. (b) X-ray rocking curve around the Bragg peak at (200) as a function of the angle $\Omega$, which corresponds to an arc along a direction nearly perpendicular to $\theta$. One observes a very broad feature which can be adjusted to three Gaussian distributions (blue lines) of crystallite orientations. (c) Same as in (a) but for a Bragg peak centered at the (002) position. Red line is a fit to a Gaussian having a full-width at half-maximum of $\sim 0.25^{\circ}$. (d) Same as in (b) but for the peak around the (002) position, although in this case two Gaussian distributions are needed to adjust the observed broad peak.}
\end{center}
\end{figure}

We also characterized our single crystals through single-crystal and
powder x-ray diffraction measurements. Figure 3 shows two typical
powder x-ray spectrum for one \emph{as-grown} BRST crystal (black
line) with a nominal stoichiometry Fe$_{1.11}$Se$_{0.4}$Te$_{0.6}$,
and for one \emph{annealed} BRST single crystal (red line) from the
same batch. These ``powder" patterns were produced from integrating
the intensities from single-crystal x-ray diffraction measurements using an area detector.
This powder like pattern was obtained on the same single crystals used for the magnetic and transport
measurements shown throughout this manuscript, by superimposing
several of the $\theta$ to $2\theta$ scans along the different
crystallographic orientations. As clearly seen, for both crystals we
observe the same set of well-defined Bragg peaks, where all peaks
can be indexed within the tetragonal unit cell having the space group symmetry \emph{P}4/\emph{nmm}. Most
importantly, we do not detect the presence of any impurity phase in
either crystal, concluding that the annealing process preserves the
original crystallographic structure and therefore the stoichiometry
of the crystals. From this first analysis it would seem that both
crystals are equivalent in structural quality despite their marked
differences in physical behavior as seen in the top panel of Fig. 2.
\begin{figure}[htb]
\begin{center}
\epsfig{file=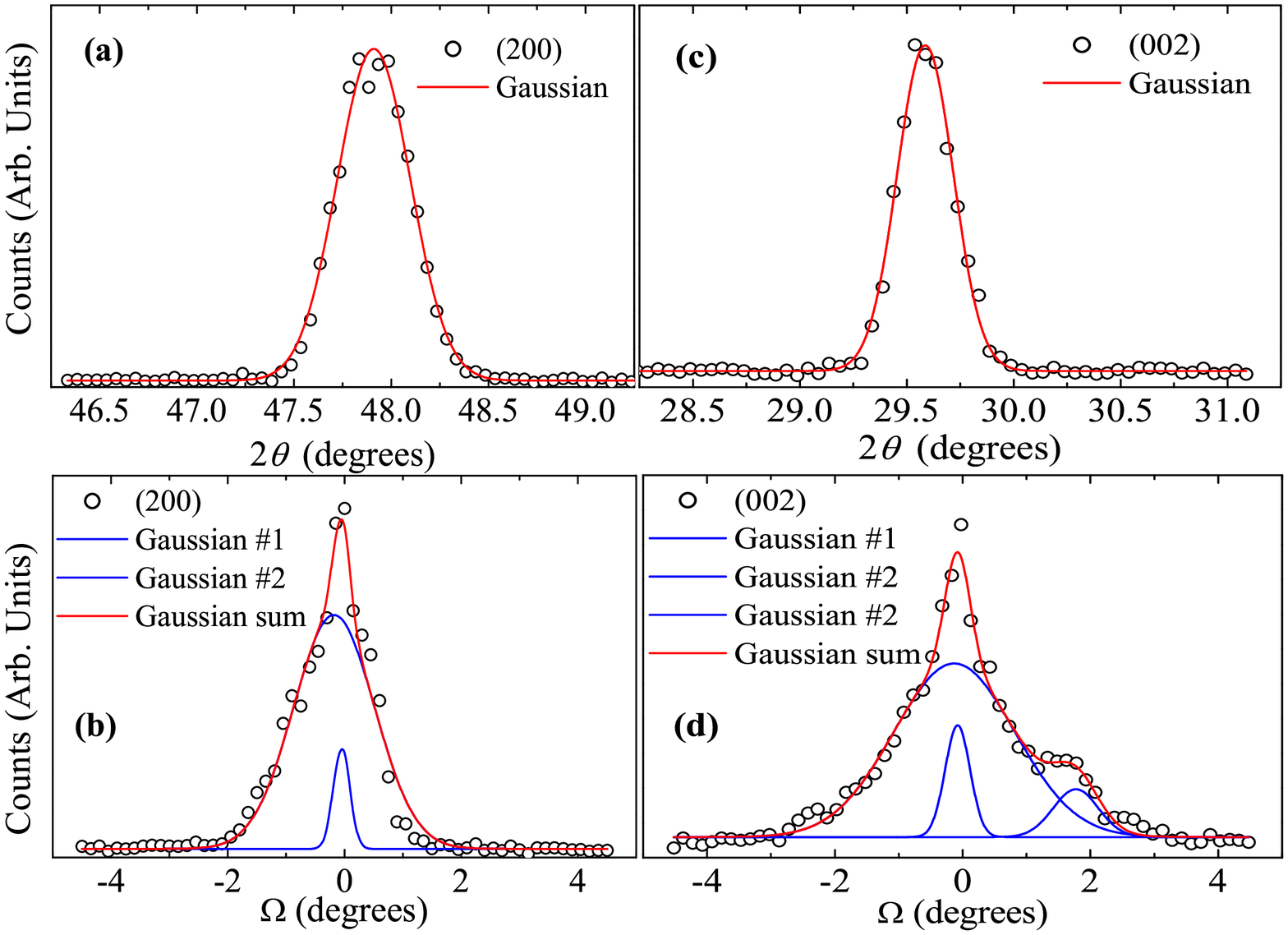, width = 8.6 cm}
\caption {(color online) (a) X-ray intensity for an \emph{annealed} BRST single-crystal of Fe$_{1+y}$Te$_{0.6}$Se$_{0.4}$ as a function of $2\theta$ for an angular scan around the Bragg-peak observed along the real-space position [\emph{abc}] = (200). Red line is a fit to a Gaussian function which yields a width at half maximum of $\sim 0.4^{\circ}$. (b) X-ray rocking curve around the Bragg peak at (200) as a function of the angle $\Omega$. One observes a very broad feature which can be adjusted to at least two Gaussian distributions (blue lines) of crystallite orientations. (c) Same as in (a) but for a Bragg peak observed along (002) position. Red line is a fit to a Gaussian which yields a full width at half-maximum of $\sim 0.3^{\circ}$. (d) Same as in (b) but for the peak around the (002) position.}
\end{center}
\end{figure}

\begin{figure}[htb]
\begin{center}
\epsfig{file=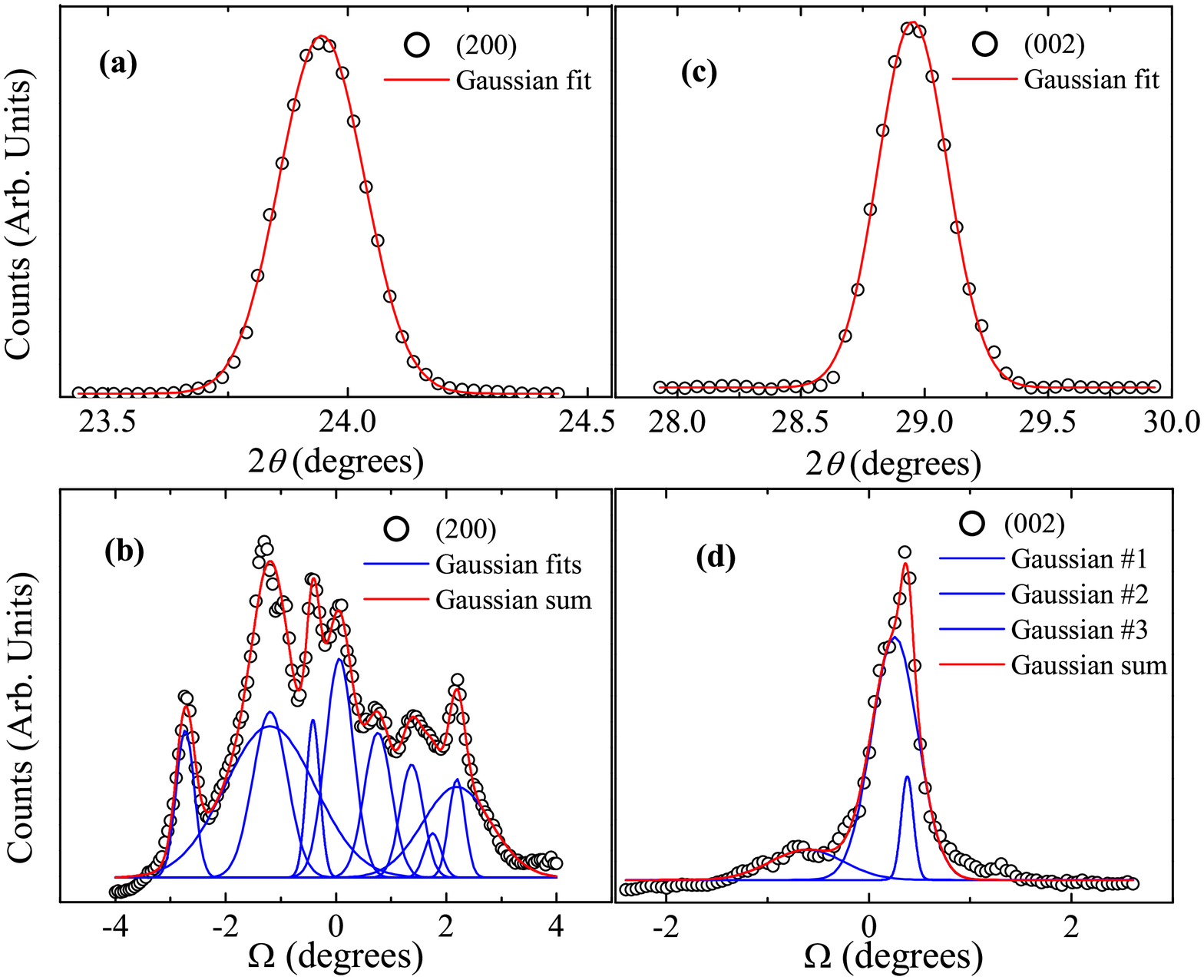, width = 8.6 cm}
\caption {(color online) (a) X-ray intensity for a single-crystal of Fe$_{1+y}$Te$_{0.55}$Se$_{0.45}$ grown by the traveling floating zone method as a function of $2\theta$ for an angular scan around the Bragg-peak observed along the (200) position. Red line is a fit to a Gaussian function which yields a width at half maximum of $\sim 0.2^{\circ}$. (b) X-ray rocking curve around the Bragg peak at (200) as a function of the angle $\Omega$. One observes a very broad feature which can be adjusted to a superposition of least ten Gaussian distributions (blue lines) of crystallite orientations. (b) Same as in (a) but for a Bragg peak observed along (002) position. Red line is a fit to a Gaussian which yields a full width at half-maximum of $\sim 0.3^{\circ}$. (d) Same as in (b) but for the peak around the (002) position.}
\end{center}
\end{figure}

To clarify if disorder and mosaicity is at the origin of these
differences in physical behavior, we performed a detailed x-ray
rocking curve analysis in both single crystals around specific Bragg
peaks. Figure 4 (a) shows the (200) Bragg reflection. By fitting the intensity as a function of
$2\theta$ to a Gaussian distribution, one obtains a full width at
half maximum (FWHM) of $\sim 0.5^{\circ}$, suggesting at first
glance a modest but sizeable mosaic spread among crystallites.
However, an exploration of the width of this peak along an arc
$\Omega$ whose orientation is nearly perpendicular to $2 \theta$
reveals a very broad feature as seen in Fig. 4 (b). This broad
maximum can be adjusted (red line) to 3 Gaussian distributions (blue
lines), with one of them showing a FWHM as large as $\gtrsim
2.5^{\circ}$. Each Gaussian would represent an ensemble or
distribution of crystallites having a similar relative orientation.
The observation of several distributions with varying widths
indicates a large degree of mosaicity around the (200) position, or
within the planes. A similar study around the (002) position shown
respectively in Figs. 4 (c) and 4 (d), reveals seemingly a smaller
degree of mosaicity among crystallites within distinct planes.

To clarify if the annealing process affects the structural degrees
of freedom, for instance, by releasing strain induced in the
single-crystals by the synthesis process, we have attempted a
similar detailed structural analysis in an annealed single crystal
of Fe$_{1+y}$Te$_{0.6}$Se$_{0.4}$. Figure 5 (a) shows the intensity
of the Bragg peak located along the (200) direction as a function of
the angle $2\theta$, while Fig. 5 (b) shows the intensity of the
same peak as function of the angle $\Omega$. The first measurement
leads to a Gaussian peak (red line) having a FWHM of $\simeq
0.4^{\circ}$ while the second one can be adjusted to two Gaussian
distributions (blue lines) with the widest distribution having a
FWHM of $\simeq 1.6^{\circ}$. Fig. 5 (c) shows the intensity of the
Bragg peak observed as the angle $2\theta$ is scanned around the
(002) direction. One extracts a FWHM of $\simeq 0.3^{\circ}$ by
fitting its intensity to a Gaussian distribution (red line).
Finally, Fig. 5 (d) shows the intensity of the Bragg peak observed
at the (002) position as a function of the angle $\Omega$. As shown
in the figure, to reproduce the observed broad feature one would
need to fit it to at least three Gaussian distributions with the
widest one having a FWHM of $\gtrsim 2.1^{\circ}$.

From this detailed comparison between the as-grown and the annealed
BRST Fe$_{1+y}$Te$_{0.6}$Se$_{0.4}$ single crystals, we can firmly
conclude that the annealing process has no or perhaps just a
marginal effect on the degree of crystallinity of the BRST grown
samples. Although as seen in Fig. 2 (a), the behavior of the
metallic state, and even the width of the superconducting transition
is affected by this treatment and as we shall see below, it also
affects $H_{c2}(T)$. In Figs. 6 (a) through (d) we show a similar rocking curve analysis
for a typical TSFZ Fe$_{1+y}$Te$_{0.55}$Se$_{0.45}$ single-crystal, and as can clearly
be seen in Figs. 6 (b) and (c) the degree of mosaicity in this single crystal, is higher
(spans over $6^{\circ} $) than what we found for the BRST crystals. Nevertheless, the width of the superconducting
transition (see Fig. 1) is considerably sharper than the width of the transition in as grown BRST crystals (see Fig. 2(a))
Therefore, from our study we can state that the mosaicity, or the relative orientation between stacked planes
in a single-crystal, has virtually no effect on the superconducting
properties (such as the width of the transition) of the 11 series.

Nevertheless, by comparing the EDX data taken on several annealed
and non-annealed BRST-grown single-crystals, on approximately a dozen
of collection sites in each crystal, one notices marked differences
between annealed and non-annealed crystals: in average the
non-annealed crystals displayed a variability $\Delta y$ of $\sim
7.7$ \% in the fraction $y$ of interstitial Fe and a $\Delta x
\simeq 8.8$ \% in the fraction $x$ of Se, when compared respectively
to $\Delta y = 1.8$ \% and $\Delta x = 6.1$ \% for the annealed
samples ($\Delta y$ is the average of the fluctuations with respect to the average value of $y$). EDX indicates that the annealing process has virtually no
effect on the average value of either $x$ or $y$, so it just
homogenizes both values (specially $y$) throughout the sample.

However, this observation does not address the origin of for
instance, the ``non-metallic" like resistivity observed in the
TSFZ-grown crystals when compared to the metallic response seen in
annealed BRST-grown crystals. For the first ones, we were able to
observe clear Bragg peaks and proceed with single-crystal x-ray
refinements in shards taken from them, indicating relatively large
values for $y$ ranging from 0.07 to 0.12. The crystallographic information file for
one of our single-crystals is available as supplemental material \cite{cif_file}.  We also performed EDX analysis on the TSFZ single crystals, finding similar amounts of excess interstitial Fe, and typical values for $\Delta y \simeq 2$ \%. Remarkably, for the second ones and despite the physical characterization shown above suggesting relatively high-quality single-crystals, we were unable to extract well-defined integrated intensities to proceed with single-crystal
x-ray refinements. For these crystals, EDX indicates nearly stoichiometric values for
$y$, i.e. $y \sim 0$ although, as previously mentioned, it tends to
underestimate the content of Fe when compared to single-crystal
x-ray refinements. These observations indicate that the physical
properties of the 11 Fe-chalcogenides are nearly oblivious to their degree
of mosaicity but depend on the amount of interstitial Fe and on
how evenly distributed it is throughout the sample.

\section{Superconducting phase diagram at high fields}

In this section we compare the high-field superconducting
phase-diagrams of several crystals synthesized by either the
floating zone or the Bridgman-Stockbarger technique. In each case,
only the crystals displaying the sharpest superconducting
transitions $\Delta T_c = T(\rho(0.9 T_c)- T(\rho(0.1T_c)) \lesssim
1.0$ K (where $\rho(0.9T_c)$ and $\rho(0.1T_c)$ are the values of
the resistivity at respectively, the onset of the resistive
transition or 90 \% of $\rho$, and the foot or 10 \% of $\rho$ at
the transition) were chosen for the electrical transport
measurements under high fields. As indicated by Fig. 2, for the
crystals grown by BRST the method, only the annealed samples showed
the optimum $T_c$ and the sharpest superconducting transitions among
all of the measured samples.
\begin{figure}[htb]
\begin{center}
\epsfig{file=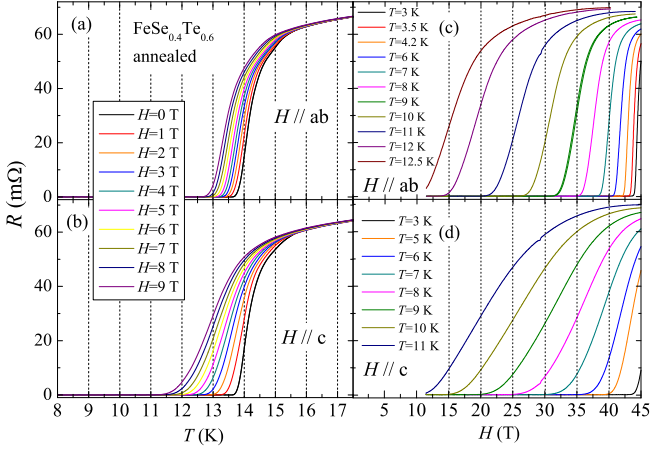, width = 8.6 cm}
\caption {(color online) Left panel: Resistance as a function of temperature $T$ across the superconducting transition for an annealed Fe$_{1.11}$Se$_{0.4}$Te$_{0.6}$ single crystal (nominal stoichiometry) synthesized by the Bridgeman technique, and respectively for fields along an in-plane direction (a) and the inter-plane c-axis (b). Right panel: Resistance as a function of field $H$ for the same single-crystal and for different temperatures, and respectively for an in-plane direction (c) and the inter-plane c-axis (d).}
\end{center}
\end{figure}

Figures 7 (a) and (b) show the resistance as a function of
temperature for an annealed Fe$_{1+y}$Te$_{0.6}$Se$_{0.4}$  single
crystal for different fields ${\bf H}\|ab$ and ${\bf H}\|c$,
respectively. Clearly, fields along the \emph{c}-axis produce
broader transitions as a function of temperature. Figures 7 (c) and
(d) show the resistance as a function of magnetic fields ${\bf
H}\|ab$ and ${\bf H}\|c$ (up to $H = 45$ T) for the same single
crystal for different temperatures. The transitions become
progressively sharper in field as the temperature decreases,
consistent with the reduction of vortex fluctuations at lower $T$.
We have also collected similar data-sets for several single-crystals
grown by the TSFZ-method with different Se and Fe contents but the
corresponding raw data is not shown here. The stoichiometric
fractions of Fe and Se were determined through a detailed
single-crystal x-ray diffraction refinement of small shards cleaved
off from each single-crystal. For these TSFZ samples the measured
superconducting transition temperatures ranged from $T_c \sim 12$ K
($y = 0.26$, $x$ or fraction of Se = 0.16) to the optimal $T_c$
value $\gtrsim 14$ K ($y = 0.06$ and $x=0.44$).
\begin{figure}[htb]
\begin{center}
\epsfig{file=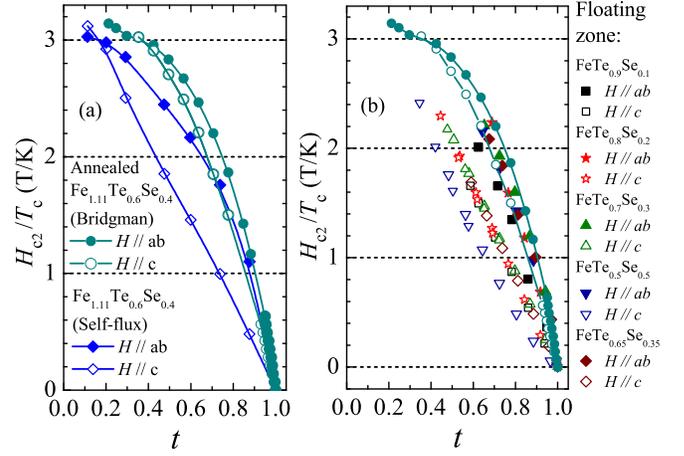, width = 8.6 cm}
\caption {(color online) (a) Comparative superconducting phase-diagram under high magnetic fields where the upper critical field $H_{c2}$ normalized by the superconducting transition temperature $T_c$, is plotted as a function of the reduced temperature $t = T/T_c$ for two single crystals: respectively for the Fe$_{1+\delta}$Se$_{0.4}$Te$_{0.6}$ single crystal in Ref. \onlinecite{fang2} and for our annealed Fe$_{1.11}$Se$_{0.4}$Te$_{0.6}$ (nominal stoichiometry) BRST single-crystal. (b) Superconducting phase-diagram for several TSFZ grown single-crystals, where the phase-diagram for our annealed Fe$_{1.11}$Se$_{0.4}$Te$_{0.6}$ BRST single-crystal is also included. Open and solid markers depict the phase-boundary between metallic and superconducting states for fields along the \emph{c}-axis and along a planar direction, respectively. All points in both graphs were extracted from the middle point of the resistive transition.}
\end{center}
\end{figure}

In Fig. 8 (a) we compare the superconducting phase-diagram at
high-fields for our annealed BRST Fe$_{1+y}$Te$_{0.6}$Se$_{0.4}$
single-crystal (circles) with the results of Ref. \onlinecite{fang2}
(squares), and in Fig. 8 (b) with the phase-diagram obtained for
several TSFZ single-crystals (various markers excluding circles). 
Here, all points correspond to the middle point of the resistive transition.
In either graph, both the field- and the temperature-axis are re-scaled
with the zero-field $T_c$ of each sample (corresponding to the
middle point of the zero-field resistive transition).
\begin{figure}[htb]
\begin{center}
\epsfig{file=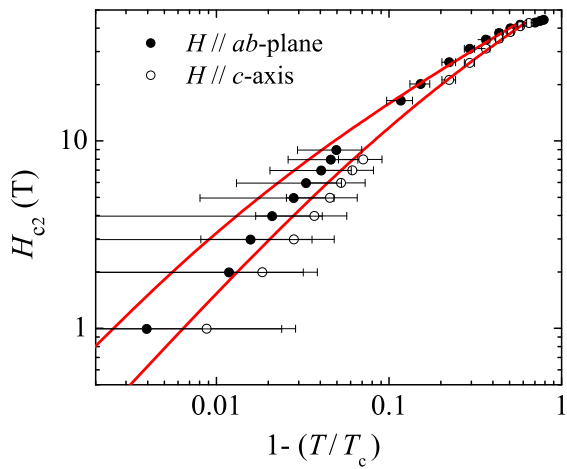, width = 7.4 cm}
\caption {(color online) Upper critical fields $H_{c2}$ as a function of $1-t$ where $t=T/T_c$ is the reduced temperature for an annealed Bridgman-Stockbarger grown single-crystal.  Red lines are fits to Eq. 1.
It yields values for the coherence lengths $\xi_{ab}$ and $\xi_{c}$ which are in good agreement with those of Ref. \onlinecite{klein}.}
\end{center}
\end{figure}
Here, we use this scaling for comparative purposes only, and not to
indicate that these compounds are placed within the dirty limit.
The plot in Fig. 7 (a) leads to a similar scaling quality
when compared with a $H_{c2}(T)/T_c^2$ as a function of $T/T_c$ plot,
previously observed in some of the Fe arsenide compounds \cite{jaroszynski} and which was deemed to be consistent
with an effective clean limit scenario due to the small
superconducting coherence lengths of the Fe-pnictide compounds.
As clearly seen in both figures, the annealed BRST-grown crystal has
considerably larger upper-critical fields than all the other
crystals, particularly for fields along the $c$-axis and at
intermediary reduced temperatures, i.e. from $\sim 35$ to $\sim 60$
\% larger $H_{c2}$s for $T/T_c \simeq 0.5$. For $T/T_c \simeq 0.2$,
the upper-critical field still is $\sim 5$ \% larger than the value
reported in Ref. \onlinecite{fang2}. On the other hand, the
anisotropy parameter $\gamma \simeq 1.8$ for $T \rightarrow T_c$ is
slightly smaller than the values reported in Ref. \onlinecite{fang2}
($\gamma \simeq 2$ for $T \rightarrow T_c$), Ref. \onlinecite{khkim}
($\gamma \simeq 3$ for $T \rightarrow T_c$), and Ref.
\onlinecite{klein} ($\gamma \simeq 3.5-4$ for $T \rightarrow T_c$).
Given that the level of structural disorder (i.e. mosaicity) seen in
crystals showing quite broad transitions and lower $T_c$s (as-grown
BRST crystals) is very similar to the level of mosaicity seen in the
annealed crystals which display relatively sharp transitions,
optimum $T_c$, and higher $H_{c2}$, one is lead to conclude that the
superconducting phase diagram of the Fe$_{1+y}$Te$_{1-x}$Se$_{x}$
series for a given value of $x$ is mostly controlled by $\Delta y$
and $\Delta x$ respectively the overall fluctuation on the value of
$y$ the fraction of interstitial Fe and in the value of $x$ the
fraction of Se.

Another intriguing observation is that the superconducting
phase-diagram for the more Fe stoichiometric samples shows a
pronounced downward curvature as $T$ is lowered which was attributed
by several authors to the Pauli limiting effect \cite{klein,
petrovic}, in contrast with for example, the linear dependence seen
in Ref. \onlinecite{fang2}) for $H_{c2}^{c}$. Indeed, all
$H_{c2}/T_c \simeq 3$ ratios shown in Fig. 7 are above the BCS
paramagnetic limit, $H_{c2}/T_c = 1.84$, indicating that the Pauli
pairbreaking is very essential.

\subsection{Two-band analysis of the $H_{c2}$ data}

In order to evaluate the contributions of both orbital and Pauli
pairbreaking effects for either field orientation, and  to compare
the results with those extracted from heat capacity measurements, we
first analyze our $H_{c2}(T)$ data at temperatures close to $T_c$
where the Ginzburg-Landau theory yields Ref.
\cite{klein,Gurevich2010}:
    \begin{equation}
    \left( \frac{H}{H_p}\right)^2 + \frac{H}{H_\text{o}} = 1 - \frac{T}{T_c}
    \label{gl}
    \end{equation}
Very close to the critical temperature, $(T_c-T)/T_c \ll
(H_p/H_\text{o})^2$, the first paramagnetic term in the left hand
side is negligible and Eq. (\ref{gl}) yields the orbital linear GL
temperature dependence, $H_{c2}=H_\text{o}(1-T/T_c)$.  At lower
temperatures, $(T_c-T)/T_c > (H_p/H_\text{o})^2$, the Pauli limiting
field $H_p$ dominates the shape of $H_{c2}(T) \propto (1-t)^{1/2} $
even in the GL domain if $H_p < H_\text{o}$. The latter inequality
is equivalent to the condition that the Maki parameter $\alpha_M
\sim H_\text{o}/H_p >1$ is large enough, assuring that the
paramagnetic effects are essential.  Shown in Fig. 9 are the log-log
plot of our $H_{c2}(T)$ as a function of $1-T/T_c$ where the red
lines are fits to Eq. (\ref{gl}). These fits are excellent for the
high-field region (excluding the highest fields), but less so very
close to $T_c$, probably due to the relatively large errors in
determining temperatures ($\Delta T \sim 25$ mK) which are inherent
to transport measurements, or perhaps due to broadening of the
resistive transition due to local $T_c$ inhomogeneities. The fit
yields $H_p^{\|c} = (72.3 \pm 3.5)$ T and $H_p^{\|ab} = (62.6 \pm
2.4)$ T for the Pauli limiting field, respectively for fields along
the \emph{c}-axis and along the \emph{ab}-plane, and
$H_{\text{o}}^{\|c} = (160.1 \pm 17.3)$ T and $H_\text{o}^{\|ab} =
(439 \pm 136)$ T for the orbital limiting fields which are in very
good agreement with those reported in Ref. \onlinecite{klein}.
Defining the effective GL coherence lengths,
$\xi_{ab}(T)=\xi_{ab}(1-T/T_c)^{-1/2}$ and
$\xi_{c}(T)=\xi_{c}(1-T/T_c)^{-1/2}$, we obtain a rather small
in-plane value of $\xi_{ab} = (\phi_0/2\pi
H_{\text{o}}^{\|c})^{1/2}= 14 \pm 4$ {\AA} and an even smaller
c-axis value of $\xi_c =\xi_{ab}H_{\text{o}}^{\|c}/H_{\text{o}}^{\|ab} \sim 5.1$ {\AA},
which is shorter than the inter-planar distance $c \sim 6$ {\AA}
\cite{tegel}, in agreement with  Ref. \onlinecite{klein}.

\begin{figure}[htb]
\begin{center}
\epsfig{file=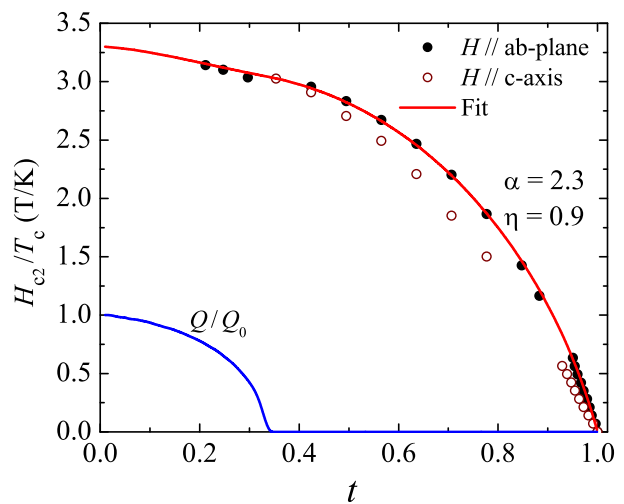, width = 8 cm}
\caption {(color online) High-field superconducting phase-diagram or upper critical-field normalized by the superconducting transition temperature at zero field as function of the reduced temperature, for an annealed BRST-grown Fe$_{1+\delta}$Se$_{0.4}$Te$_{0.6}$ single crystal, where the red line is a fit to the model described in the text. Blue line depicts the temperature dependence of the modulation wave-vector for a FFLO state, which would explain the shape of the phase-boundary at lower temperatures.}
\end{center}
\end{figure}

The phenomenological Eq. (\ref{gl}) only describes the range
$T\approx T_c$, but does not take into account the possibility of
the FFLO state which occurs if $\alpha_M>\alpha_c$ is large enough
($\alpha_c \approx 1.8$ for a single parabolic band \cite{fflo}). To
analyze our $H_{c2}(T)$ data at all temperatures and to reveal the
microscopic meaning of the scaling fields $H_p$ and $H_{\text{o}}$,
we use a two-band generalization of the WHH theory
\cite{Gurevich2010} in the clean limit, taking into account both
orbital and paramagnetic pair-breaking, and the possibility of the
FFLO with the wave vector $Q(T,H)$. In this case the equation for
$H_{c2}$ is given by
    \begin{gather}
    a_1G_1+a_2G_2+G_1G_2=0,
    \label{mgh} \\
    G_1=\ln t+2e^{q^{2}}\operatorname{Re}\sum_{n=0}^{\infty}\int_{q}^{\infty}due^{-u^{2}}\times
    \nonumber \\
    \left[\frac{u}{n+1/2}-\frac{t}{\sqrt{b}}\tan^{-1}\left(
    \frac{u\sqrt{b}}{t(n+1/2)+i\alpha b}\right)\right]G_2=
    \label{U1}
    \nonumber \\
    \ln t+2e^{sq^{2}}\operatorname{Re}\sum_{n=0}^{\infty}\int_{q\sqrt{s}}^{\infty}due^{-u^{2}}\times
    \nonumber \\
    \left[\frac{u}{n+1/2}-\frac{t}{\sqrt{b\eta}}\tan^{-1}\left(
    \frac{u\sqrt{b\eta}}{t(n+1/2)+i\alpha b}\right)\right]
    \label{U2}
    \end{gather}
The FFLO wave vector $Q(T,H)$ is determined self-consistently by the
condition that the solution $H_{c2}(T,Q)$ of Eq. (\ref{mgh}) is
maximum, $a_1=(\lambda_0 +\lambda_{-})/2w$,
$a_2=(\lambda_0-\lambda_{-})/2w$,
$\lambda_{-}=\lambda_{11}-\lambda_{22}$,
$\lambda_0=(\lambda_{-}^2+4\lambda_{12}\lambda_{21})^{1/2}$,
$w=\lambda_{11}\lambda_{22}-\lambda_{12}\lambda_{21}$, $t=T/T_c$,
and
    \begin{gather}
    b=\frac{\hbar^{2}v^{2}_1 H }{8\pi\phi_{0}k_B^2T_c^2g_1^2},\qquad\alpha=\frac{4\mu\phi_{0}g_1k_BT_{c}}{\hbar^{2}v^2_1}=\frac{\pi k_B T_c m}{E_F m_0},
    \label{parm1} \\
    q^{2}=Q_{z}^{2}\phi_{0}\epsilon_1/2\pi H, \qquad \eta = v_2^2/v_1^2, \qquad
    s=\epsilon_2/\epsilon_1.
    \label{parm2}
    \end{gather}
Here $v_l$ is the in-plane Fermi velocity in band $l={1,2}$,
$\epsilon_{l}=m_{l}^{ab}/m_{l}^c$ is the mass anisotropy ratio,
$\mu$ is the magnetic moment of a quasiparticle, $\lambda_{11}$ and
$\lambda_{22}$ are the intraband pairing constants, and
$\lambda_{12}$ and $\lambda_{21}$ are the interband pairing
constants, and $\alpha\approx 0.56\alpha_M$.  The factors
$g_1=1+\lambda_{11}+|\lambda_{12}|$ and
$g_2=1+\lambda_{22}+|\lambda_{21}|$ describe the strong coupling
Eliashberg corrections. For the sake of simplicity, we consider here
the case of $\epsilon_1=\epsilon_2=\epsilon$ for which $s=1$, and
$H_{ab}$ is defined by Eqs.~ (\ref{mgh}) and (3) with $g_1=g_2$ and
rescaled $q\to q\epsilon^{-3/4}$, $\alpha\to\alpha\epsilon^{-1/2} $
and $\sqrt{b}\to \epsilon^{1/4}\sqrt{b}$ in $G_1$ and $\sqrt{\eta
b}\to \epsilon^{1/4}\sqrt{\eta b}$ in $G_2$ \cite{Gurevich2010}.

Equation (\ref{mgh}), which describes $H_{c2}(T)$ at all
temperatures, simplifies close to $T_c$ where it reduces to Eq.
(\ref{gl}) with
    \begin{gather}
    H_{\text{o}}=\frac{48\pi\phi_0k_B^2T_c^2}{7\zeta(3)\hbar^2[v_1^2+v_2^2 + (v_1^2-v_2^2)\lambda_{-}/w]}
    \label{ho} \\
    H_p=\frac{2\pi k_BT_c}{\mu\sqrt{7\zeta(3)}}
    \label{hp}
    \end{gather}
where $\zeta(3)\approx 1.202$. The slope $H_{c2}' = |dH_{c2}/dT|$ is
maximum at $T_c$ where Eqs. (2) and (\ref{ho}) yield
  \begin{equation}
  H_{c2}'=\frac{24\pi\phi_0k_B^2T_c}{7\zeta(3)\hbar^2(c_{+}v_1^2+c_{-}v_2^2)}
  \label{slope}
  \end{equation}
For identical bands, ($v_1=v_2=v$, $\lambda_{11}=\lambda_{22}$,
$c_+=c_-=1/2$), Eq. (\ref{slope}) reduces to the single-band GL
expression, $H_{c2}'=\phi_0/2\pi T_c\xi_0^2$, where $\xi_0=(\hbar
v/\pi k_BT_c)[7\zeta(3)/48]^{1/2}$ is the GL coherence length in the
clean limit \cite{Gurevich2010}. For the $s^\pm$ pairing
$(c_{+}\rightarrow c_{-}\rightarrow 1/2)$, the dependence of
$H_{c2}'$ on the materials parameters resembles the behavior of
$H_{c2}(T)$ in the $s^{++}$ dirty limit: for strong band asymmetry
($\eta\ll 1$ or $\eta\gg 1$),  $H_{c2}$ in Eq. (\ref{slope}) is
limited by the band with {\it larger} Fermi velocity, similar to
$H_{c2}$ mostly limited by the band with larger diffusivity for the
$s^{++}$ case \cite{ag}. Paramagnetic effects decrease the slope of
$H_{c2}'$ and reduce the effect of band asymmetry.

Figure 10 shows the fit of the measured $H_{c2}(T)$ to
Eq.~(\ref{mgh}) for $H\|ab$. For this field orientation, the
resistive transitions in Fig. 7 (c) are considerably sharper than for
$H\|c$. This allowed us to clearly define the middle point of the
resistive transition for the traces taken at lower temperatures,
increasing the number of the $H_{c2}$ data points. For the sake of
simplicity, we consider here the case of
$\epsilon_1=\epsilon_2=\epsilon $ for which
$\gamma=\epsilon^{-1/2}$, and $H_{ab}$ is defined by Eqs.~
(\ref{mgh}) and (3) with rescaled $q\to q\epsilon^{-3/4}$,
$\alpha\to \alpha\epsilon^{-1/2} $ and $\sqrt{b}\to
\epsilon^{1/4}\sqrt{b}$ in $G_1$ and $\sqrt{\eta b}\to
\epsilon^{1/4}\sqrt{\eta b}$ in $G_2$ \cite{Gurevich2010}. The fit
in Fig. 10 was done for $s^{\pm}$ pairing with
$\lambda_{11}=\lambda_{22}=0$, $\lambda_{12}\lambda_{21}=0.25$,
$\eta = 0.9$, $H\| ab$, $\alpha=2.3$. Equation (\ref{mgh}) describes
$H_{c2}^c (T)$, $H_{c2}^{ab}(T)$ and $\gamma_H
(T)=b_\|(T)/\sqrt{\epsilon}b_\bot(T)$ where $b_\|(T)$ is the
solutions of Eq.~(\ref{mgh}) for $H \|c$  very well.

To see if the observed values of $H_{c2}'$ are at least
qualitatively consistent with Eq. (\ref{slope}) and the materials
parameters of chalcogenides, we use the ARPES data
\cite{arpes2,arpes3,arpes4} which give  $E_F\simeq   20-50$ meV.
Taking $E_F = 30$ meV in Eq. (\ref{slope}) yields $H'^{ab}_{c2}$
[T/K] $\simeq 0.64\gamma(m/m_0)$, so the observed
$H'^{ab}_{c2}=H_o^{ab}/T_c=31.4$ T/K and $\gamma =
H_o^{ab}/H_o^{c}\approx 2.75$ corresponds to the value of $m\simeq
15m_0$ consistent with the ARPES data of $m/m_0$ = 3-20 for
different FS pockets \cite{arpes2}. These estimates show that the
paramagnetic effects for Fe-11 chalcogenides are indeed essential,
and the parameter $\alpha^{ab}=\gamma\alpha$ defined by Eq.
(\ref{parm1}) is greater than 1. For $E_F= 30$ meV and $T_c=14$ K,
we obtain $\alpha^{ab} \simeq 0.13\gamma m/m_0$, giving
$\alpha^{ab}\sim 5$.  The values of $\alpha>1$ indicate that
paramagnetic pair-breaking becomes so strong that it can trigger the
FFLO instability at lower temperatures.

As follows from the results shown in Fig.~10, the anisotropy
parameter $\gamma_H(T)$ decreases as $T$ decreases. This behavior
reflects the significant role of the Zeeman pair-breaking in FeSeTe
given that $\alpha_\| = \alpha/\sqrt{\epsilon}=2.3$ for $H \bot c$
is well above the single-band FFLO instability threshold,
$\alpha\approx 1$ \cite{Gurevich2010}. In this case $\gamma_H (T)$
near $T_c$ is determined by the orbital pairbreaking and the mass
anisotropy $\epsilon$, but as $T$ decreases, the contribution of the
isotropic Zeeman pairbreaking increases, resulting in the decrease
of $\gamma_H(T)$.

The fit to our experimental data based on the solution of
Eq.~(\ref{mgh}) predicts the FFLO transition for $H ||ab$. In this
case Fig. 10 shows that the FFLO wave vector $Q(T)$ appears
spontaneously at $T< T_F\approx 0.35T_c\approx 5$K, corresponding in optimally doped samples to magnetic fields $H \gtrsim 42 $ T.
Here $Q(T)$ increases from zero at $T=T_F$ to the maximum $Q_0$ at $T=0$, where
$Q_0=4\pi T_c b(0)^{1/2}/\hbar v_1$. In addition to the strong Pauli
pairbreaking $\alpha > 1$, the FFLO transition requires weak
impurity scattering in the clean limit $\ell \gg \xi$ where $\ell$
is the mean free path due to elastic scattering on impurities
\cite{fflo}. The latter condition is likely satisfied in the
crystals studied in this work, given the very short coherence
lengths $\xi_{ab}\sim 14$ {\AA} and $\xi_c\sim 5$ {\AA} extracted from the
fit described above. Thus, the chalcogenides are very good
candidates to study the FFLO transition by magneto-transport,
specific heat, magnetic torque or NMR measurements.

\section{Analysis of the critical region}

Recent heat-capacity measurements on FeSe$_{0.5}$Te$_{0.5}$ under field \cite{serafin}
found  that it strongly resembles behavior previously observed in the much more anisotropic high-$T_c$
cuprates, where strong fluctuation effects have been found to wipe out the phase
transition at $H_{c2}$. In FeSe$_{0.5}$Te$_{0.5}$ these fluctuations were found to be strongly anisotropic \cite{serafin}.
Here, we proceed to study the critical regime preceding the superconducting transition in our annealed BRST-grown single-crystals.
In the critical region of the metal to superconducting transition,
and in the limit of strong magnetic fields, one expects a scaling
form for the thermodynamic functions. At very high fields if the
quasiparticles are confined within the lowest Landau level, the
transport of charge carriers becomes nearly one-dimensional along
the direction of the applied field. Fluctuation effects close to the
superconducting transition are expected to be enhanced by the lower
effective dimensionality of the system. In the critical regime the
fluctuation conductivity was calculated in Ref. \onlinecite{ullah}
by including a quartic term in the free energy within the Hartree
approximation, and obtained a scaling law for the fluctuation
conductivity $\Delta \sigma$ in magnetic fields, in terms of
unspecified scaling functions $F_{\text{2D}}$ and $F_{\text{3D}}$,
valid for two-dimensional and three-dimensional superconductors,
\cite{ullah} respectively:
\begin{equation}
\Delta \sigma (H)_{\text{2D}} = \left ( \frac{T}{H} \right)^{1/2}
F_{\text{2D}} \left( \alpha \frac{T-T_c(H)}{\sqrt{TH}}\right),
\label{2d}
\end{equation}
\begin{equation}
\Delta \sigma (H)_{\text{3D}} = \left ( \frac{T^2}{H} \right)^{1/3}
F_{\text{3D}} \left( \beta \frac{T-T_c(H)}{(TH)^{2/3}}\right),
\label{3d}
\end{equation}
Where $\alpha$ and $\beta$ are characteristic constants of a
specific material. Presumably, these functionals remain valid even
if the quasiparticles are not confined to just the lowest Landau
level, but are placed in a few higher Landau levels, as long as the
interaction between quasiparticles in distinct Landau levels remains
negligible. \cite{zlatko, zlatko1} In the cuprates, either type of
scaling has been observed in materials having distinct degrees of
electronic/superconducting anisotropies. \cite{cuprates}

We define the contribution of the fluctuations to the conductivity
$\Delta \sigma$ as the difference between the normal state
conductivity $\sigma_n = 1/ \rho_n$ ($\rho_n$ is the normal state
resistivity) and the measured conductivity $\sigma = 1/ \rho$. Here
$\sigma_n$  is obtained from a polynomial fit of the resistivity in
a temperature interval $\Delta T \sim 2T_c  $ above $T_c$, where the
contribution of the fluctuations to $\sigma$ should become
negligible. Here, $T_c$ is defined as the middle point of the resistive transition, or the 50 \%
\begin{figure*}[htb]
\begin{center}
\epsfig{file=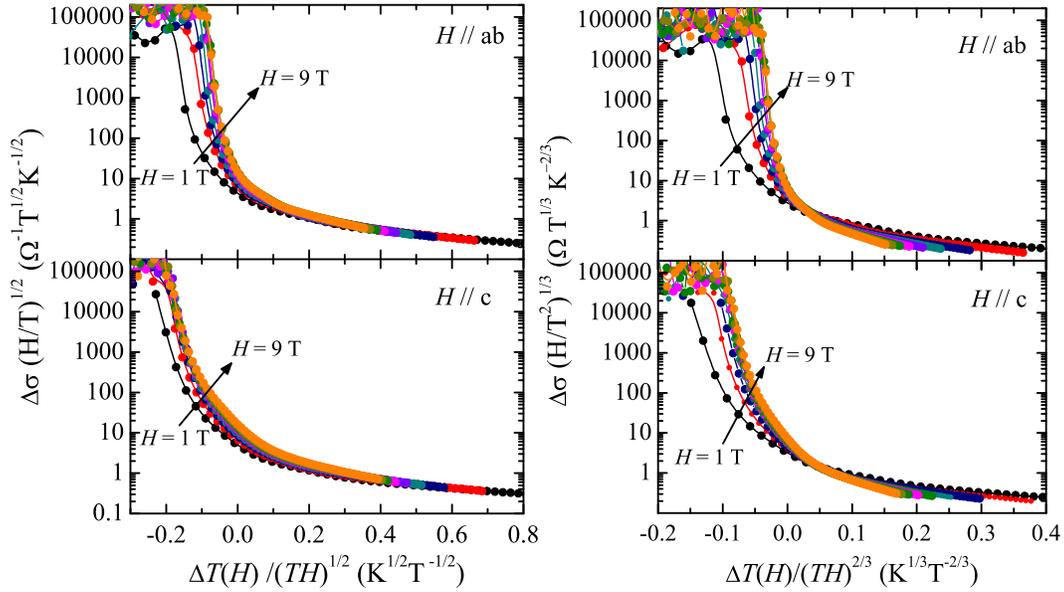, width = 14 cm}
\caption {(color online) Left panel: $\Delta \sigma (H/T)^{1/2}$ versus $\Delta T (H) /(TH)^{1/2}$ for the annealed BRST single-crystal and for both field orientations, as expected for the two-dimensional scaling relation of Eq. (2). Here, $\Delta \sigma = (1 / \rho - 1 / \rho_n)$ where $1/ \rho$ is the resistivity across the superconducting transition and $1/ \rho_n$ is the resistivity in the normal state preceding the transition (see main text), while $\Delta T (H)= [T-T_c(H)]$. Right Panel: $\Delta \sigma H^{1/3}/T^{2/3}$ versus $[T-T_c(H)] /(TH)^{2/3}$ as expected for the three-dimensional scaling relation in Eq. (3).}
\end{center}
\end{figure*}

In the left panel of Fig. 11, we plot $\Delta \sigma (H/T)^{1/2}$
versus $[T-T_c(H)] /(TH)^{1/2}$ for the annealed BRST single-crystal
for both field orientations, as expected for the two-dimensional
scaling relation in Eq.~(\ref{2d}). For fields above a relatively
small value of just 4 T, and for both orientations, this
two-dimensional scaling clearly succeeds in collapsing all the
curves over an extended temperature range particularly above $T_c$.
We checked that this statement is still valid if the onset of the resistive
transition (90 \% of the resistance of the normal state) is used as the criteria to define $T_c$.
Given that the high field resistive transitions shown in Fig. 7 are
remarkably broad, it is not possible to unambiguously define the
functional form of $\rho_n(H)$ in order to also collapse those
curves in Fig. 11. To compare this two-dimensional scaling relation
with the three-dimensional one, in the right panel of Fig. 11, we
plot $\Delta \sigma H^{1/3}/T^{2/3}$ versus $[T-T_c(H)] /(TH)^{2/3}$
as expected for the three-dimensional case described by Eq.~
(\ref{3d}). Clearly, the two-dimensional relationship provides a
better scaling of the data both above and below $T_c$. Indeed, in
the three-dimensional case, the curves at different fields open like
a fan above $T_c$ and also tend to separate from each other below
this critical temperature. On the other hand, in the 2D case, the
curves above 4 T basically overlap over the entire range. The 2D
behavior of fluctuation conductivity, already observed in the Fe
arsenides, \cite{pallecchi,marina} implies that the inter-plane
coherence length $\xi_c$ is smaller or in the order of the
inter-plane spacing as found above, or, equivalently, that the coupling
between adjacent superconducting planes is smaller than the
intra-plane condensation energy. Notice, that our scaling analysis is consistent
with the report of very anisotropic superconducting fluctuations according to  
heat capacity measurements \cite{serafin}. Two-dimensional scaling also
suggests that such fluctuations contribute to the broad
transitions seen in Fig. 7, which become considerably sharper at
lower temperatures.

\section{discussion}

There are several puzzling aspects in the Fe$_{1+y}$Te$_{1-x}$Se$_x$
family that have yet to be understood and/or conciliated. For
instance, optical spectroscopy measurements in
Fe$_{1.06}$Te$_{0.88}$S$_{0.14}$ indicates the absence of a Drude
peak in optical conductivity, suggesting the absence of
well-defined coherent quasi-particles in this material
\cite{optical}. Although, for FeSe$_{0.42}$Te$_{0.58}$ the in-plane optical conductivity
is found to be indeed describable within a Drude-Lorentz model \cite{homes} but with an incoherent
response along the inter-planar direction \cite{moon}.
Therefore, one could even ask if the concept of
Fermi surface would be applicable to these compounds. In sharp
contrast, a recent angle resolved photoemission study on
FeSe$_{0.42}$Te$_{0.58}$ \cite{tamai} reveals a Fermi surface which
is in relative good qualitative agreement with the generic Fermi
surface proposed for the family of Fe pnictides.
Namely, composed by two concentric hole-like cylindrical surfaces at
the $\Gamma$ point and two electron-like cylinders at the \emph{M}
point of the Brillouin zone, which are characterized by
a large, FS-sheet dependent, effective-mass enhancement.
This indicates that correlations are particularly relevant for these materials which is
consistent with the size of the anomaly observed in the heat
capacity at the superconducting transition \cite{maohc} or the large
$H_{c2}$ reported here and by other groups \cite{fang2,khkim,klein}.
This geometry would be consistent with the observation of a
resonance mode characterized by a wave-vector connecting both
types of cylindrical Fermi surfaces \cite{qiu} pointing towards an
itinerant, nesting-like mechanism for the origin of the
spin-fluctuations in these systems. Although, this
scenario seems difficult to conciliate with the rather
large magnetic moment of 1.6 to 1.8 $\mu_B$ for the non-interstitial
Fe(I) ion as extracted from careful magnetic susceptibility
measurements in \emph{superconducting} samples \cite{yoshimura}, and
does not explain the  observation of short-range magnetic-order in
(superconducting samples) for the same wave-vector as the one
extracted for the magnetically ordered state of Fe$_{1+y}$Te \cite{bao}.
The interstitial Fe(II) is presumably characterized by an even larger moment of $\sim$ 2.5 $\mu_B$
\cite{bao, yoshimura} and from the superconducting perspective it
should act as a magnetic impurity.

A recent DFT + DMFT study suggests that a combination of Hund's rule
and structural degrees of freedom, such as the pnictogen height and
the bond angle between Fe and the pnictide element, are the key
parameters defining the role of correlations, i.e. the degree of
localization and therefore the concomitant size of the magnetic
moment of Fe in iron pnictide compounds \cite{kotliar}. For
instance, the larger the size of the pnictogen atom or equivalently
the larger the distance between Fe and this element, or the greater
the deviation with respect to the ideal tetrahedral angle of
109.5$^{\circ}$, the larger the degree of localization in the Fe
site. Or equivalently, the stronger the correlations on those Fermi
surface sheets having a marked $t_{2g}$ character which would favor
the development of a magnetic instability. It is therefore easy to
speculate on the role played by the interstitial Fe(II): it should
favor local lattice distortions that are likely to further distort
the Fe-pnictogen angle away from 109.5$^{\circ}$ favoring local
magnetism in detriment of superconductivity. The larger the content
in Fe(II) the stronger this detrimental effect, while larger
fluctuations in $y$ could create randomly distributed nearly
magnetic patches having a stronger Fe(II) content which could lead
to pair-breaking effects. This rather simple scenario would explain
why the annealed BRST-grown samples display clearer and sharper
superconducting transitions when compared to the non-annealed ones
despite a similarly poor crystallinity.

As for the presence of the FFLO state \cite{fflo}, it is presumably sensitive
to impurities being difficult to conciliate with the presence of magnetic Fe(II)
atoms. However, our results suggest that our BRST-grown samples (annealed
as well as non-annealed) are closer to having a nearly
stoichiometric Fe content. Furthermore, the
low-$T$ upturn in $H_{c2}^c(T)$ indicative of the FFLO instability
as observed in this work, was also observed by other groups, see for example, Ref.
\onlinecite{khkim}. Finally, the interplay between magnetism and
the FFLO-state is still poorly understood \cite{kenzelmann},
so its existence in the Fe$_{1+y}$Te$_{1-x}$Se$_x$ will only be
clarified by future calorimetric, magnetic torque or NMR
measurements at high fields.

Finally, given that the inter-plane coherence length $\xi_c$ is in the order of the inter-planar distance
it is relatively easy to understand why the mosaic spread associated with relative
orientation among Fe[Se,Te] planes has little effect on the superconducting properties of these materials.
Although the superconducting anisotropy is small, the size of $\xi_c$ implies that the Cooper pairs are virtually localized
within the planes thus remaining almost oblivious to the potential pair-breaking effects imposed by the grain boundaries located between the planes.

\section{Conclusion}

In conclusion, we find that the Fe[Te,Se] planes are well stacked along the c-axis, forming
domains with well aligned planes spanning a certain coherence length along the c-axis. These domains
display a small twist to each other, in the order of a degree or less relative, defining grain boundaries.
Our study strongly indicates that these domain boundaries have little or no effect on the superconducting
properties of the Fe$_{1+y}$Te$_{1-x}$Se$_{x}$ series.
However, a relatively weak disorder in the distribution of interstitial Fe is
found to be detrimental to the superconducting properties of the
Fe$_{1+y}$Te$_{1-x}$Se$_x$ series, broadening the
width of the superconducting transition and suppressing the
diamagnetic response. We attribute this to local
lattice distortions associated with the interstitial Fe(II) which
are likely to locally increase the electronic correlations leading
to a higher degree of carrier localization favoring magnetism in
detriment of superconductivity. In samples displaying the sharpest superconducting
transitions and a very clear diamagnetic signal, we observe a mild
upturn in the upper critical field at lower temperatures consistent with
the Fulde-Ferrel-Larkin-Ovchinnikov instability. These
samples display metallic resistivity suggesting also the formation
of coherent quasi-particles at lower temperatures in contrast to
previous reports \cite{optical}, and implying that a relatively mild
excess of randomly distributed Fe(II) atoms may be enough to
suppress the phase coherence. Finally, our estimates  of the
inter-plane coherence length $\xi_c \sim 5.1 < 6$ {\AA} where $c
\simeq$ 6 {\AA} is the inter-planar lattice constant, is consistent
with the quasi-two-dimensional superconducting
fluctuations in the critical region (also observed by other groups\cite{pallecchi,marina}).

\section{Acknowledgements}
The NHMFL is supported by NSF through NSF-DMR-0084173 and the
State of Florida.  L.~B. is supported by DOE-BES through award DE-SC0002613.
JW, TG and TS acknowledge support from FSU.

\end{document}